\documentclass[aps,prx,showpacs,superscriptaddress,nofootinbib,notitlepage,twocolumn]{revtex4-1}
\usepackage{graphicx}
\usepackage{color}
\usepackage{transparent}
\usepackage{amsmath,amssymb,amsfonts,amssymb}
\usepackage{epstopdf}
\usepackage{float}
\usepackage{ragged2e}
\usepackage{tikz}
\def\checkmark{\color{green}\tikz\fill[scale=0.4](0,.35) -- (.25,0) -- (1,.7) -- (.25,.15) -- cycle;\color{black}}

\newcommand{\bra}[1]{\ensuremath{\left\langle#1\right|}}
\newcommand{\ket}[1]{\ensuremath{\left|#1\right\rangle}}

\newcommand{\tr}[0]{\text{tr}}

\begin{document}




\title[]{Experimental Snapshot Verification of non-Markovianity with Unknown System-Probe Coupling}




\author{Henri Lyyra}\email{henri.s.lyyra@jyu.fi}
\affiliation{Department of Physics and Nanoscience Center, University of Jyv\"askyl\"a, FI-40014 University of Jyv\"askyl\"a, Finland}
\email{henri.s.lyyra@jyu.fi}

\author{Olli Siltanen}
\affiliation{Laboratory of Quantum Optics, Department of Physics and Astronomy, University of Turku, FI-20014, Turun yliopisto, Finland}
\affiliation{Turku Centre for Quantum Physics, Department of Physics and Astronomy, University of Turku, FI-20014, Turun yliopisto, Finland}

\author{Jyrki Piilo}
\affiliation{Laboratory of Quantum Optics, Department of Physics and Astronomy, University of Turku, FI-20014, Turun yliopisto, Finland}
\affiliation{Turku Centre for Quantum Physics, Department of Physics and Astronomy, University of Turku, FI-20014, Turun yliopisto, Finland}

\author{Subhashish Banerjee}
\affiliation{Indian Institute of Technology Jodhpur, Jodhpur 342011, India}

\author{Tom Kuusela}
\affiliation{Laboratory of Quantum Optics, Department of Physics and Astronomy, University of Turku, FI-20014, Turun yliopisto, Finland}
\affiliation{Turku Centre for Quantum Physics, Department of Physics and Astronomy, University of Turku, FI-20014, Turun yliopisto, Finland}

\begin{abstract}
We apply the recently proposed quantum probing protocols with an unknown system-probe coupling to probe the convex coefficients in mixtures of commuting states. By using two reference states instead of one as originally suggested, we are able to probe both lower and upper bounds for the convex coefficient. We perform extensive analysis for the roles of the parameters characterizing the double peaked Gaussian frequency spectrum in the Markovian-to-non-Markovian transition of the polarization dynamics of a single photon. We apply the probing of the convex coefficient to the transition-inducing frequency parameter and show that the non-Markovianity of the polarization dynamics can be confirmed with a single snapshot measurement of the polarization qubit performed at unknown time and even with unknown coupling. We also show how the protocol can identify Markovian and non-Markovian time intervals in the dynamics. The results are validated with single photon experiments.
\end{abstract}

\maketitle

\section{Introduction}

Due to its foundational role in realistic quantum systems and the implementability of quantum technologies, the study of open quantum systems has attracted a lot of attention \cite{petru,RHbook,banerjeebook}. Whenever a quantum system interacts with some other system, its environment, the system and environment form together a closed system whose dynamics is unitary. Due to this interaction, the dynamics of the reduced system state is not necessarily unitary and it is said to be an open system. When the open system is used to store or process information, the open system dynamics causes information flow from the system to the environment and correlations between the system and environment. This means that all of the information initially encoded in the open system cannot be retrieved just by measurements on the evolved open system. 

Luckily, the loss of information is not always monotonic and in some cases the information has been shown to partially return to the open system as the interaction is prolonged. Multiple definitions of quantum non-Markovianity based on such revivals of information and also other dynamical properties have been proposed and there is no agreement on a single definition  \cite{BLP,lorenzomeasure,BCM,hemeasure,RHPreview,BLPreview,vegareview,lihallreview,EPLN-MWhatIs}. In this paper we refer to non-Markovianity as revivals of the trace distance of the optimal pair of initial states, namely the BLP non-Markovianity, which directly quantifies the increases of distinguishability of the state pair during the open system's dynamics \cite{BLP}. Since for our specific dynamics 
BLP non-Markovianity is equivalent to multiple definitions of non-Markovianity, such as CP divisibility, our results apply also more generally \cite{nonMarkEquivalence,teittinen2018}.

In addition to fundamental interest of the nature of information flow, 
non-Markovianity has found many uses in making quantum protocols more efficient under noisy circumstances \cite{EPLN-MGoodFor}. Non-Markovianity has been experimentally shown to increase the success rate of Deutsch-Josza algorithm implemented in NV-center in diamond
, and allow for perfect superdense coding and quantum teleportation with mixed two photon polarization states in noisy transmission \cite{panalgorithm,karlssonEPL,laine2014,liu2020}. 
Non-Markovianity has been shown to help in entanglement generation 
and distribution
, and it has been shown to improve the secure key rate in quantum key distribution \cite{huelgarivasplenio2012,Xiang2014,utagi}. 
Multiple different measures and indicators of non-Markovianity have been developed but directly experimentally confirming non-Markovianity requires comparison of the system state at two different times, such that the information measure in question is larger at the later time. 

Even though the open system dynamics is generally harmful for the information carrier, the information flow into the combined system-environment state can be exploited in specific types of measurements. Since the initial state of the environment influences how the open system evolves in time, measurements on the evolved system can be used to deduce some unknown properties of the environment. Such measurements are useful in situations where direct measurements on the system could hinder the operation of a quantum device - or in the worst case - destroy its information carriers. In these so-called quantum probing measurements, the unitary coupling operator describing the total system-environment dynamics is commonly assumed as known. This assumption is used to form a mapping between the measured open system acting as the probe, and the unknown properties of the environment, which is the system of interest in the measurement \cite{probing2006,probingnonmark2013,probing2013,probing2016,probing2017,haikka}. While using such assumption has proven successful in specific cases, it has certain disadvantages: As the whole measurement strategy is based on one fixed unitary coupling, it has to be faithfully implemented in the experiment. If the coupling is changed, the protocol fails as the connection between the unknown properties and the evolved probe state changes also.

Recently, a new approach to quantum probing was introduced in \cite{tukiainen}. The approach is based on a generalized data processing inequality of the so-called $\alpha$-fidelities which was shown useful for multiple purposes. Among its applications a quantum probing protocol with unknown system-probe coupling was proposed. In the protocol, the generalized data processing inequality is applied to compare two probe states after they have interacted with the system prepared in an unknown state and some reference state. This comparison yields to information of the unknown state. 

Later, such protocol was experimentally implemented in a single photon experiment \cite{lyyrasiltanen2020}. In that paper, the upper bound for the width of a Gaussian frequency spectrum was probed by measurements on a polarization probe qubit that had interacted with frequency in a combination of quartz plates rotated in randomly chosen angles, corresponding to an unknown system-probe coupling. These works showed that it is possible to construct and implement measurements whose action is arbitrarily chosen and completely unknown but still result to non-trivial and reliable information.

In this paper, we show how the aforementioned probing protocols based on the generalized data processing inequalities can be used to extract lower and upper bounds for the convex coefficients in mixtures of commuting states by using unknown system-probe coupling. We apply this result in detecting the global feature of non-Markovianity of the dynamics and identifying the Markovian and non-Markovian time intervals from snapshot measurements at unknown time - and even with completely unknown system-probe coupling. Thus we show that the probing protocol can be used to obtain qualitative information about the probe's important dynamical features in addition to the static properties of the system.

Previously, a method for determining the non-Markovianity in terms of the minimal deviation of a snapshot channel from the set of all dynamical semigroups and CP-divisible dynamical maps was developed \cite{WolfEisertCirac}. In contrast, our goal is to determine if the whole dynamical map is non-Markovian and to identify the Markovian and non-Markovian time intervals of the dynamics in terms of information back flow as revivals of trace distance - all from a single snapshot at unknown time and with unknown system-probe coupling. 
We note here, that a sequence of system state preparations can also be used to define and study non-Markovianity in another way by combining them with a sequence of control operations performed during the dynamics \cite{pollock2018}. 

The paper is structured as follows: First, we discuss the necessary background of open quantum systems, quantum probing, and generalized data processing inequalities. Then, we develop the measurement strategy for probing the convex coefficients in mixtures of commuting states with unknown system-probe couplings and list some of its possible applications. After that, we concentrate on a specific quantum optical implementation, namely probing the convex coefficient of a mixture of two Gaussian frequency spectra. We analyze extensively the roles of the parameters of the double peaked spectrum in the non-Markovianity of the polarization dynamics and show that by using our probing measurements, it is possible to verify the non-Markovianity of the polarization dynamics with no assumptions on the system-probe coupling or any of the parameters of the dynamics-inducing frequency state. Finally, the protocol is experimentally realized in an all-optical setup and the results are discussed.



\section{Open Quantum Systems, Quantum Probing, and Generalized Data Processing Inequalities}\label{theory}

We say that a quantum system $A$ is open if it interacts with some other system, the environment $B$. Commonly, it is assumed that $A$ and $B$ are uncorrelated before the interaction begins. The dynamics of the total closed system $AB$ is described by a unitary coupling $U$ which makes $A$ and $B$ interact. The state of the system $A$ after the interaction becomes
\begin{equation}\label{reduced}
\Phi(\rho) = \text{tr}_B [ U ( \rho \otimes \xi ) U^\dagger ]\,,
\end{equation}
where $\rho$ and $\xi$ are the initial states of $A$ and $B$, respectively, and tr$_B [X]$ is the partial trace of $X$ over the Hilbert space of the environment $B$. As a concatenation of completely positive and trace preserving (CPTP) maps $\Phi$ is also CPTP, or in other words a \emph{channel}. 

The effects of CPTP maps have been widely studied and it has been shown that, in terms of some information measures, information can never increase in channels. One of such commonly used measures is the \textit{trace distance} of states $\rho_1$ and $\rho_2$, defined as
\begin{equation}
D_{tr} (\rho_1,\rho_2) := \frac{1}{2} \text{tr} \left[ \sqrt{ (\rho_1-\rho_2)^\dagger (\rho_1-\rho_2) } \right]\,,
\end{equation}
where $\sqrt{X}$ is the unique positive operator satisfying $\sqrt{X}\sqrt{X} = X$ for $X \ge 0$. Trace distance gives directly the probability of distinguishing two equally likely states $\rho_1$ and $\rho_2$ in a single realization of an optimal measurement. Trace distance was shown to satisfy the following \textit{data processing inequality}
\begin{equation}
D_{tr} (\rho_1,\rho_2) \ge D_{tr} \big(\Phi( \rho_1 ), \Phi( \rho_2 ) \big)\,,
\end{equation}
for any channel $\Phi$, and any pair of initial states $\rho_1$ and $\rho_2$, meaning that the distinguishability of any pair of states can never increase in CPTP maps \cite{nielsenchuang}. Also the commonly used \textit{fidelity of quantum states}, defined as
\begin{equation}
F_{1/2} (\rho_1, \rho_2) := \tr \left[ \sqrt{ \sqrt{\rho_2} \, \rho_1 \, \sqrt{\rho_2} } \right]\,.
\end{equation}
satisfies the following data processing inequality
\begin{equation}
F_{1/2} (\rho_1, \rho_2) \le F_{1/2} \big( \Phi( \rho_1 ), \Phi( \rho_2 ) \big)\,,
\end{equation}
which means that the similarity of two states in terms of fidelity can never decrease in CPTP maps.

Even though the data processing inequalities say that information can be only lost from the open system $A$, it still remains in the state of the total closed system AB whose dynamics is unitary. This feature has been exploited in the so-called \textit{quantum probing} protocols. In quantum probing, the system $S$ has some unknown parameter $x$ and the goal is to evaluate the value of $x$ without making direct measurements on $S$. This can be the case for example   when $S$ is a part of a device such as a quantum computer and we want to monitor its behavior while it is running, but direct measurements would disturb or even destroy it. 

In quantum probing, direct measurements are avoided by preparing a disposable probe system $P$ in some known state $\rho$ and letting it interact with $S$ under some unitary $U$. Then, measurements on $P$ are used to extract information on $x$. In the above description of open quantum system, probe $P$ corresponds to the open system $A$, and system $S$ is its environment $B$. Equation \eqref{reduced} shows that the channel $\Phi$ that is induced on probe $P$ depends on the initial state $\xi$ of $S$. In the usual quantum probing protocols, the unitary coupling $U$ is known and a mapping between the unknown parameter $x$ in $\xi$ and the transformed probe state $\Phi(\rho)$ is used to evaluate $x$ from measurements on $\Phi(\rho)$. Such protocols depend on using a specific coupling $U$, and if it is not known or cannot be properly implemented, the unknown parameter of $S$ cannot be mapped to the evolved probe state $\Phi(\rho)$ and thus the protocol cannot be applied. 

\begin{figure}[H]
\centering
  \includegraphics[width=0.8\linewidth]{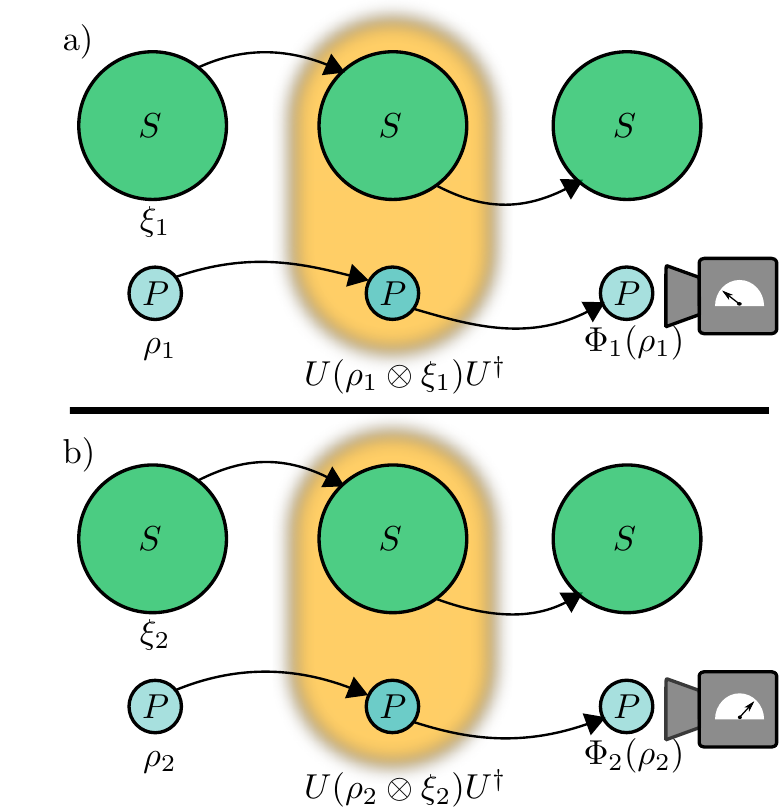}
  \caption{
  \textbf{The unknown coupling quantum probing approach.} The system $S$ interacts with the probe $P$. The unitary $U$ is the same in panels a) and b). The analytical form of $\xi_1$ and $\xi_2$ is known, but some parameters in $\xi_1$ and $\xi_2$ are different. Thus, the induced probe channels $\Phi_1$ and $\Phi_2$ may be different. The system-probe coupling $U$ is unknown, and consequently so are the solutions for the channels $\Phi_1$ and $\Phi_2$. Nevertheless, comparing the measured probe states $\Phi_1(\rho_1)$ and  $\Phi_2(\rho_2)$ can be used to gain reliable and non-trivial information on the unknown parameter.
  }
  \label{env_in_dyn-12}
\end{figure}

To see how quantum probing can be performed even if $U$ is unknown, let us consider the two cases in Fig.~\ref{env_in_dyn-12}. Once the coupling $U$ is fixed, the dynamics of the probe $P$ depends on the initial state $\xi$ of the system $S$. If $S$ is prepared in different states $\xi_1$ and $\xi_2$ and it is coupled to $P$, this can induce two different channels $\Phi_1$ and $\Phi_2$ on $P$ even if $U$ is the same for both initial states of $S$. 

This observation was combined with the data processing inequality to form a mathematical tool for studying open quantum systems. This tool utilizes the comparisons between the initial system and probe states, and the evolved probe states \cite{tukiainen}. The {\it $\alpha$-fidelity of states} was defined for $\alpha \in (0,1)$ as \cite{tukiainen}
\begin{eqnarray}
&&F_{\alpha} \big(\rho_1, \rho_2\big) := \tr \left[ \left(\rho_2^{\frac{1-\alpha}{2\alpha}} \rho_1 \, \rho_2^{\frac{1-\alpha}{2\alpha}} \right)^\alpha \right]\,.
\end{eqnarray}
We note that in the special case $\alpha=1/2$, $F_{1/2}$ becomes the commonly used fidelity of states.

Now, let us consider the $\alpha$-fidelities in the context of Fig.~\ref{env_in_dyn-12}. The unitary coupling $U$ between $P$ and $S$ is fixed but in Fig.~\ref{env_in_dyn-12} a) and b) the initial states of $P$ and $S$ can be different. Thus, different choices of states $\xi_1$ and $\xi_2$ of $S$ induce channels $\Phi_1$ and $\Phi_2$ to $P$ in the interaction, respectively. 
In this open system picture, it was shown that the $\alpha$-fidelities satisfy the following \textit{generalized data processing inequality} \cite{tukiainen}
\begin{eqnarray}\label{eq:mainineq2}
F_{\alpha} \big(\rho_1, \rho_2\big) F_{\alpha} \big(\xi_1, \xi_2\big) \leq F_{\alpha} \big(\Phi_1(\rho_1), \Phi_2(\rho_2)\big)
\,,
\end{eqnarray}  
$\forall\,\alpha\in[1/2,1)$ and for all unitary couplings $U$ and initial states $\rho_1$, $\rho_2$, $\xi_1$, and $\xi_2$.


Equation \eqref{eq:mainineq2} allows us to develop new kind of probing protocols which do not require any knowledge of the system-probe coupling. Let us assume that the analytical form of $F_{\alpha} \big(\xi_1, \xi_2\big)$ is calculated and we want to get bounds for some parameters characterizing the system state $\xi_1$ or $\xi_2$. In the protocol, the experimenter lets the system, prepared in states $\xi_1$ and $\xi_2$, interact with probes prepared in known states $\rho_1$ and $\rho_2$, respectively, as in Fig.~\ref{env_in_dyn-12}. Then, the evolved probe states $\Phi_1(\rho_1)$ and $\Phi_2(\rho_2)$ are determined with tomographical measurements. Solving for the unknown parameter in Eq.~\eqref{eq:mainineq2} and inserting the measured density operators in $F_{\alpha} \big(\Phi_1(\rho_1), \Phi_2(\rho_2)\big)$ yields non-trivial bounds for the unknown parameter.

By using the subadditivity w.r.t.~tensor products \cite{wilde}, unitary invariance \cite{nielsenchuang}, and the data processing inequality, trace distance can be shown to satisfy the following generalized data processing inequality in the open quantum system picture of Fig.~\ref{env_in_dyn-12} \cite{mikkothesis}
\begin{equation}\label{tracedistanceineq}
D_{tr} \left( \Phi_1(\rho_1),\Phi_2(\rho_2) \right) \leq D_{tr} \left( \rho_1, \rho_2 \right) + D_{tr} \left( \xi_1, \xi_2 \right)\,,
\end{equation}
for all unitary couplings $U$ and initial states $\rho_1$, $\rho_2$, $\xi_1$, and $\xi_2$. As described above for $\alpha$-fidelities, also the generalized data processing inequality of trace distance can be used to construct quantum probing protocols with unknown system-probe couplings.

The generalized data processing inequalities \eqref{eq:mainineq2} and \eqref{tracedistanceineq} are our main mathematical tool in the probing protocols. In the next section we use them to construct quantum probing protocols with unknown system-probe couplings to determine lower and upper bounds for convex coefficients in mixtures of commuting states which we apply later in snapshot verification of the probe's non-Markovianity.

\section{Probing of Convex Coefficients}


Let us consider a state of interest, given by $\xi_1 = p \xi_2 + (1 - p)\xi_3$, which is a convex combination of two known commuting states $\xi_2$ and $\xi_3$ with the unknown convex coefficient $p \in [0,1]$. We prepare three probe systems in the states $\rho_1$, $\rho_2$, $\rho_3$ and let them interact with the system in the state $\xi_1$ and the two reference states $\xi_2$ and $\xi_3$, respectively. By measuring the evolved probe states $\Phi_1(\rho_1)$, $\Phi_2(\rho_2)$, and $\Phi_3(\rho_3)$, we can perform a probing measurement to obtain bounds for the convex coefficient $p$.

The commuting states $\xi_2$ and $\xi_3$ can be written diagonal in the same basis $\{\ket{k}\}_{k = 1}^{d^S}$ as $\xi_2 = \sum_k \lambda_k \ket{k}\bra{k}$ and $\xi_3 = \sum_k \nu_k \ket{k}\bra{k}$. Thus we get
\begin{align}
F_{\alpha} ( \xi_1, \xi_2 ) 
&= \sum_{k} \left(  [ p \lambda_k + (1 - p) \nu_k ] \right)^\alpha   \lambda_k^{1-\alpha} \\
&\geq \sum_{k} \left(  p \lambda_k \right)^\alpha   \lambda_k^{1-\alpha} 
= p^\alpha  \\
&\Rightarrow F_{\alpha} ( \xi_1, \xi_2 ) \geq p^\alpha  \,.\label{commstatefidelity}
\end{align}

More specifically, if $\xi_2$ and $\xi_3$ are orthogonal, the $\alpha$-fidelity between the state of interest $\xi_1$ and the first reference state $\xi_2$ becomes
\begin{equation}\label{statefidelity}
F_\alpha (\xi_1, \xi_2) = p^\alpha\,.
\end{equation}

By using Eq.~\eqref{commstatefidelity} or \eqref{statefidelity} in Eq.~\eqref{eq:mainineq2}, we get an upper bound for the convex coefficient as
\begin{equation}\label{ainequality}
p \leq \left( \frac{F_{\alpha} \big(\Phi_1(\rho_1), \Phi_2(\rho_2)\big)}{F_{\alpha} \big(\rho_1, \rho_2\big)} \right)^{1/\alpha}\,.
\end{equation}

Similarly, by using the $\alpha$-fidelity between the state of interest $\xi_1$ and the second reference state $\xi_3$, we get
\begin{align}
F_{\alpha} ( \xi_1, \xi_3 ) &\geq (1- p)^\alpha  \,,\label{commstatefidelity2}
\end{align}
for commuting $\xi_2$ and $\xi_3$, and
\begin{equation}\label{statefidelity2}
F_\alpha (\xi_1, \xi_3) = (1-p)^\alpha\,,
\end{equation}
when $\xi_2$ and $\xi_3$ are orthogonal. As a consequence, we get also a lower bound for the convex coefficient $p$ as
\begin{equation}\label{ainequality2}
p \geq 1 - \left( \frac{F_{\alpha} \big(\Phi_1(\rho_1), \Phi_3(\rho_3)\big)}{F_{\alpha} \big(\rho_1, \rho_3\big)} \right)^{1/\alpha}\,,
\end{equation}
whenever $\xi_1 = p \xi_2 + (1 - p)\xi_3$ where $\xi_2$ and $\xi_3$ are orthogonal or commute.

By combining Eqs.~\eqref{ainequality} and \eqref{ainequality2}, we get
\begin{gather}
1 - \left( \frac{F_{\alpha_3} \big(\Phi_1(\rho_1), \Phi_3(\rho_3)\big)}{F_{\alpha_3} \big(\rho_1, \rho_3\big)} \right)^{1/\alpha_3} \nonumber \\ 
\leq ~ p ~ \leq \label{ainequalityfull} \\
\left( \frac{F_{\alpha_2} \big(\Phi_1(\rho_1), \Phi_2(\rho_2)\big)}{F_{\alpha_2} \big(\rho_1, \rho_2\big)} \right)^{1/\alpha_2} \,, \nonumber
\end{gather}
where $\alpha_2$ and $\alpha_3$ are independent parameters in the interval $[1/2,1)$.

For the trace distance between the state of interest $\xi_1$ and the commuting reference states $\xi_2$ and $\xi_3$ we get
\begin{align}
D_{tr} (\xi_1,\xi_2) 
&= \frac{1}{2} \sum_k (1-p) |\lambda_k - \nu_k |  \\
&\leq \frac{1}{2} \sum_k (1-p) (\lambda_k + \nu_k) 
= 1-p
\\
&\Rightarrow D_{tr} (\xi_1,\xi_2)  \leq 1-p
\,,
\end{align}
where we used the same spectral decompositions for $\xi_2$ and $\xi_3$ as above, and similarly 
\begin{align}
D_{tr} (\xi_1,\xi_3) &\leq p\,.
\end{align}

If $\xi_2$ and $\xi_3$  are orthogonal, we get
\begin{align}
D_{tr} (\xi_1,\xi_2) &= 1 - p\,,\\
D_{tr} (\xi_1,\xi_3) &= p\,.
\end{align}

As a consequence of Eq.~\eqref{tracedistanceineq}, we get another set of bounds for $p$ as 
\begin{gather}
D_{tr} ( \Phi_1( \rho_1 ), \Phi_3( \rho_3 ) ) - D_{tr} ( \rho_1,\rho_3 ) \nonumber \\
~ \leq ~ p ~ \leq ~ \label{trainequality} \\
1 - \left[ D_{tr} ( \Phi_1( \rho_1 ), \Phi_2( \rho_2 ) ) - D_{tr} ( \rho_1,\rho_2 ) \right]\,, \nonumber
\end{gather}
whenever $\xi_1 = p \xi_2 + (1 - p)\xi_3$ where $\xi_2$ and $\xi_3$ are orthogonal or commute.

Interestingly, this result does not depend on knowing anything about how the system and probe interact, as it is based on the same approach as studied in \cite{tukiainen,lyyrasiltanen2020}. This means that the protocol is not sensitive to imperfections in the implementation of the system-probe coupling $U$, and the same strategy can be used for multiple different couplings to obtain even tighter bounds. As Eq.~\eqref{ainequalityfull} 
has the freedom to choose $\alpha_2$ and $\alpha_3$, the bounds corresponding to each measurement of $\Phi_1( \rho_1 ),\,\Phi_2( \rho_2 )$, and $\Phi_3( \rho_3 )$ can be optimized w.r.t.~the $\alpha$ parameters.


The bounds of the convex coefficient $p$ can be used for different purposes. First of all, when $\xi_2 \perp \xi_3$, the purity \cite{teikobook} of the state $\xi_1$ can be given as $P(\xi_1):=\text{tr}[\rho^2] = P(\xi_2)p^2 + P(\xi_3)(1 - p)^2$, so when the orthogonal states $\xi_2$ and $\xi_3$ in the convex combination are fixed, the measured bounds of $p$ can be used to get bounds also for the purity of $\xi_1$. 

Secondly, when $\xi_2 \perp \xi_3$, the von Neumann entropy \cite{nielsenchuang,teikobook} of $\xi_1$ becomes $S(\xi_1) = p S( \xi_2 ) + ( 1 - p )S( \xi_3 ) - [ p \ln (p) + (1-p) \ln (1-p)]$, so bounds of $p$ yield bounds for von Neumann entropy, as the states $\xi_2$ and $\xi_3$ are known. If $\xi_1$ represents a two-qubit state, where $\xi_2$ and $\xi_3$ are two different Bell states, the entanglement can be quantified with the concurrence measure \cite{concurrence,lyyraconcurrence} as $C(\xi_1) = |2p - 1|$. As in the case of purity and von Neumann entropy, also bounds of entanglement can be experimentally determined with our approach.

The same strategy can be used to probe upper bounds of the $N$ convex coefficients $p_i$ if the state $\xi_1$ is a mixture $\xi_1 = \sum_{i = 2}^{N+1} p_i \xi_i$ where $\xi_i$ commute. If the eigenbasis of $\xi_1$ is known, it can be written as $\xi_1 = \sum_{i = 1}^{d^S} \lambda_i \ket{\phi_i}\bra{\phi_i}$, where $\ket{\phi_i}$ is the eigenstate corresponding to the eigenvalue $\lambda_i$. As a consequence, our strategy can be used to obtain upper bounds for all the eigenvalues by performing the probing measurement with $d^S$ reference states where $d^S$ is the dimension of the Hilbert space of the system.

Next, we apply our probing protocol to a quantum optical system where the convex coefficient influences the non-Markovianity in the probe dynamics. We perform extensive analysis for the system and determine the critical values of the convex coefficient determining whether non-Markovianity appears in the probe dynamics for any combination of the other system parameters.

\section{Snapshot verification of non-Markovianity}

\begin{figure}[H]
\centering
  \includegraphics[width=0.95\linewidth]{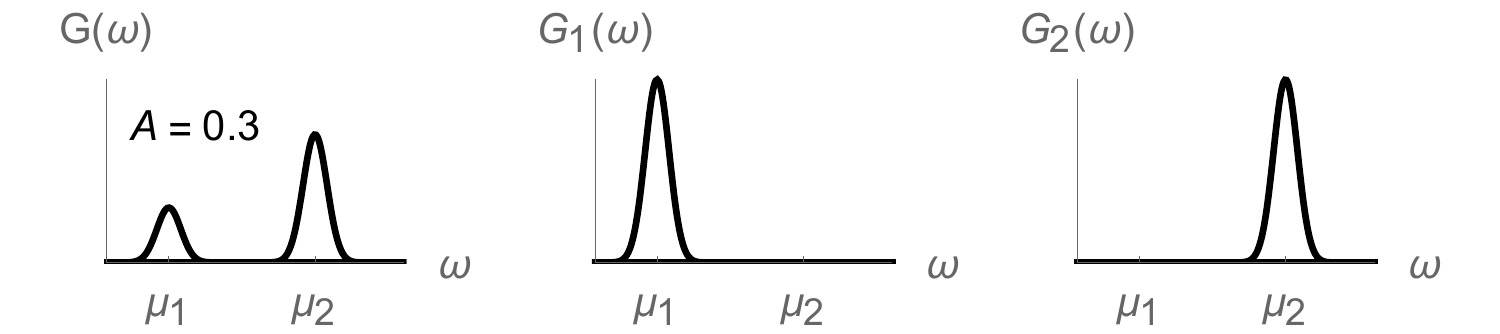}
  \caption{
  \textbf{Illustration of the frequency states $\xi_1,\,\xi_2$, and $\xi_3$.} The state of interest $\xi_1$ is a convex combination of the two reference states $\xi_2$ and $\xi_3$, and its spectrum is given by $G(\omega) = A G_1(\omega) + (1-A)G_2(\omega)$. The amplitude parameter $A$, the width of the Gaussians, and the distance of the central frequencies together control the transition of polarization probe's dynamics between Markovian and non-Markovian. 
  }
  \label{gaussians}
\end{figure}

From now on we apply the above results in a specific optical system where the convex coefficient influences whether the dynamics of the probe is Markovian or non-Markovian. Our system of interest is the frequency of a single photon and our probe is the photon's polarization. Usually in the open system literature, the photon's polarization is referred to as the system and its frequency is the environment, but since we probe the frequency by measurements on the polarization, we use the names probe and system.

Our goal is to determine from a snapshot measurement whether the polarization probe's dynamics is Markovian or non-Markovian when the frequency has been prepared in the state $\xi_1$. 
The state corresponds to a double peaked Gaussian spectrum with an unknown relative peak amplitude parameter $A$, central peak frequencies $\mu_1$ and $\mu_2$, and peak widths $\sigma$. In order to measure the properties, we prepare the frequency of two other photons into reference states $\xi_2$ and $\xi_3$, characterized by single peaked Gaussian spectra. The frequency states can be written as
\begin{align}\label{systemStates}
\begin{aligned}
&\xi_1 = A \xi_2 + (1 - A)\xi_3\,, 
~~~~~ A \in [0,1]\,,\\
&\xi_2 = \int G_1( \omega ) \ket{\omega} \bra{\omega} d \omega \,,\\
&\xi_3 = \int G_2( \omega ) \ket{\omega} \bra{\omega} d \omega \,, \\
& \text{where}~ G_k(\omega)  = \frac{1}{ \sqrt{2 \pi \sigma^2} } e^{- \frac{ (\omega - \mu_k )^2}{2 \sigma^2}}\,,~k\in \{1,2\} \,. 
\end{aligned}
\end{align}
Here, $\sigma$ is the standard deviation, and $\mu_k$ is the mean frequency of the Gaussian distribution $G_k(\omega)$, illustrated in Fig~\ref{gaussians}. 
Here,  $\omega$ are the frequency values appearing with probability  $G_k(\omega)$. Since the frequency states $\xi_2$ and $\xi_3$ commute, we can apply the quantum probing protocol to obtain lower and upper bounds for $A$ by using Eqs.~\eqref {ainequalityfull} and \eqref{trainequality}.
When the polarization and frequency interact in a birefringent medium such as a combination of quartz plates with fast axes aligned, the reduced dynamics of the polarization qubit becomes
\begin{equation}\label{dephasing}
\rho(t) = \Phi^t \big( \rho(0) \big) = 
\begin{pmatrix}
\rho_{HH} & \kappa(t)\rho_{HV}\\
\kappa^*(t)\rho_{VH} & \rho_{VV}
\end{pmatrix}\,,
\end{equation}
where $\kappa(t) = \int f(\omega)e^{i 2 \pi \omega \Delta n t } d \omega$ 
 is the decoherence function and $f(\omega)$ is the frequency spectrum \cite{setuphamiltonian}. Here $\Delta n = n_H - n_V$ is the birefringence of the medium, where $n_H$ and $n_V$ are the refractive indices in the horizontal (H) and vertical (V) directions, respectively. This polarization-frequency model has been recently popular in the studies of quantum information in open quantum systems  \cite{liu2011,setuphamiltonian,karlssonLyyra,karlssonEPL,sina2017,liu2018,sina2020,olli,liu2020,lyyrasiltanen2020,olliarxiv}. If the fast axes of the quartz plates in the combination are not aligned, the dynamics becomes significantly more complicated.

We concentrate on the non-Markovianity given by the commonly used BLP measure of non-Markovianity \cite{BLP}, which is based on the trace distance. Trace distance quantifies the distinguishability of two states, which can be interpreted as the amount of information encoded into a sequence of information carriers prepared in the two given states. Thus, increases of trace distance mean increases in information. 
If $D_{tr} (\Phi^t(\rho_1),\Phi^t(\rho_2))$ is a monotonically decreasing function of time, the states $\Phi^t(\rho_1)$ and $\Phi^t(\rho_2)$ become less and less distinguishable as time goes on, and the dynamics described by $\Phi^t$ is Markovian in terms of the BLP measure. On the other hand, if $D_{tr} (\Phi^t(\rho_1),\Phi^t(\rho_2))$ increases at some times, the distinguishability increases and thus the dynamics is non-Markovian.

Here, we note that for the pure dephasing dynamics of the form Eq.~\eqref{dephasing} the BLP non-Markovianity is equivalent to many other indicators of non-Markovianity, such as revivals of quantum channel capacity, entanglement assisted classical channel capacity \cite{BCM}, violation of CP-divisibility, Bloch volume, and $l_1$ coherence norm \cite{teittinen2018}, so our results apply directly to them too. For the pure dephasing dynamics, BLP non-Markovianity is critical for tightness of the quantum speed limit bound. Thus, our results have also implications on the optimality of the speed of the state dynamics, while the connection does not exist for the set of all qubit dynamical maps \cite{deffnerlutz,xuspeed,teittinen2019}.

For dynamics of the form Eq.~\eqref{dephasing}, the optimal pair of initial states in the BLP non-Markovianity measure can be chosen as $\rho_1^{opt} = \ket{+}\bra{+}$ and $\rho_2^{opt} = \ket{-}\bra{-}$, where $\ket{\pm} = \frac{1}{\sqrt{2}} ( \ket{H} \pm \ket{V})$ \cite{wissmankarlsson}. The dynamics is BLP non-Markovian if and only if $\frac{d}{d t} |\kappa(t)| > 0$ at some time and otherwise it is BLP Markovian. By analytically solving the decoherence function induced by the frequency in state $\xi_1$, we see that the polarization dynamics is BLP non-Markovian if and only if 
\begin{align}\label{nmcondforA}
\begin{aligned}
h(A) :=(1-A) A &>  
\frac{ 2\pi \Delta nt \sigma ^2}
{\theta ( \Delta n t, \sigma, \Delta\mu )}\,,\text{ and} \\
\theta ( \Delta n t, \sigma, \Delta\mu ) & > 0\,,
\end{aligned}
\end{align}
for some $t > 0$, where we have denoted
\begin{align}\nonumber
\theta ( \Delta n t, \sigma, \Delta\mu ) := & 4 \pi  \Delta nt \sigma ^2 - 4 \pi \Delta nt \sigma ^2 \cos \left( 2 \pi  \Delta nt \Delta\mu\right) \\
& - \Delta\mu \sin \left( 2 \pi \Delta n t \Delta\mu \right)\,,\text{ and} \\
\Delta\mu := & |\mu_2 - \mu_1|\,.
\end{align}
\begin{figure}[t!]
    \centering
    \includegraphics[width=0.95\linewidth]{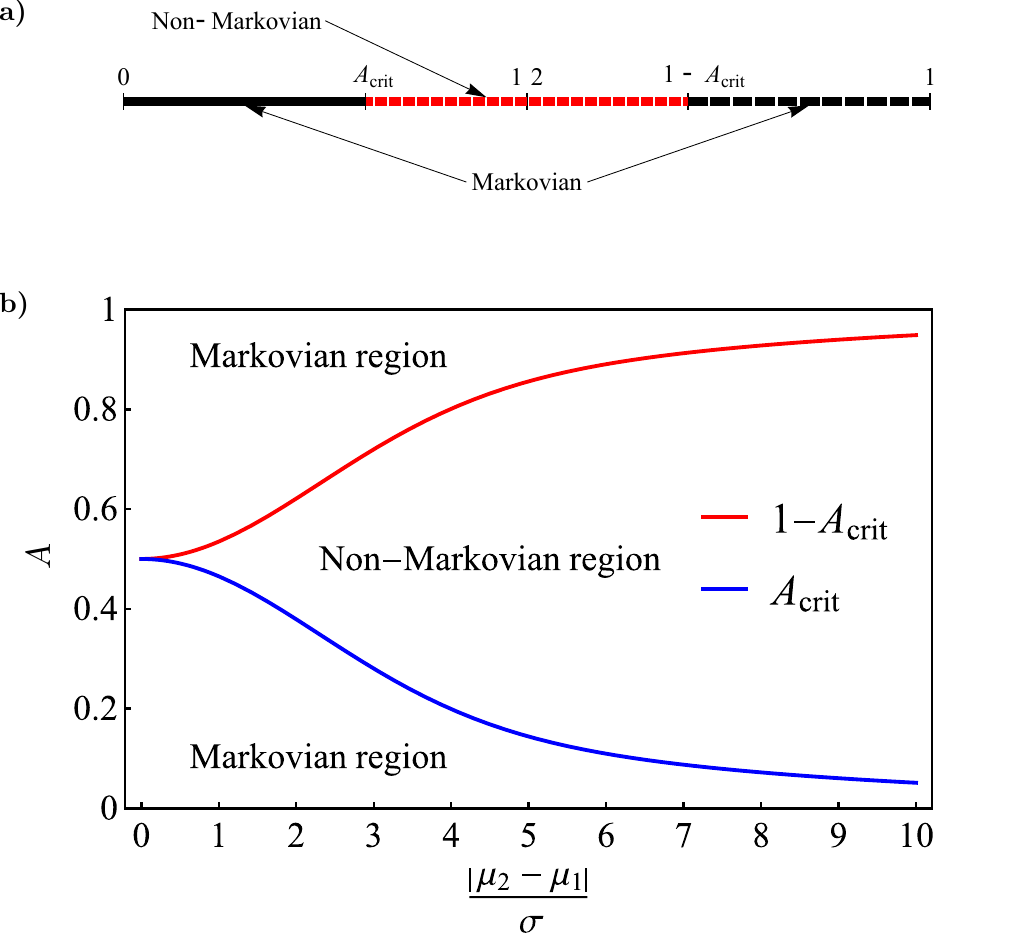}
    \caption{
  \textbf{The role of the amplitude parameter $A$ in the Markovian-to-non-Markovian transitions of the polarization dynamics.} \textbf{a)} The qubit dynamics is non-Markovian if $A \in [A_{crit} , 1 - A_{crit} ]$ and Markovian otherwise. The value of $A_{crit}$ depends on the other parameters of the double-peaked Gaussian spectrum. \textbf{b)} The boundary values between Markovian and non-Markovian regions as a function of the rescaled distance of the Gaussian peaks $|\mu_2 - \mu_1| / \sigma$.
  }\label{acrit}
\end{figure}

\begin{figure*}[t!]
\begin{minipage}[t!]{0.8\textwidth}
\centering
  \includegraphics[width=0.8\textwidth]{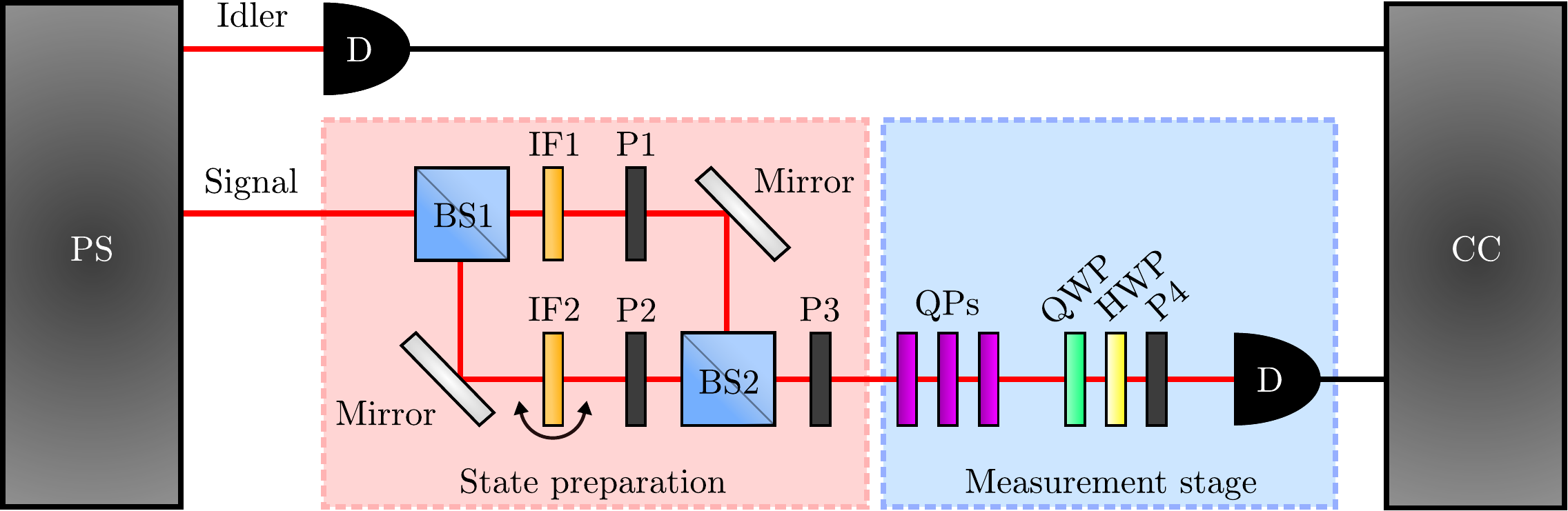}
\end{minipage}
  \caption{
  \textbf{The experimental setup.} The photon source PS produces a pair of photons. The idler photon arrives at single photon detector D and triggers the coincidence counter CC to expect the signal photon. First, the frequency-polarization state of the signal photon is prepared in the state preparation stage. The photon goes through a beamsplitter BS1 allowing to manipulate the photon independently in two distinct spatial paths. Interference filters IF1 and IF2 in their own branches are used to filter the frequency spectrum to narrower Gaussians $G_1$ and $G_2$. IF2 can be tilted to shift the central frequency of the Gaussian spectrum in the lower branch. Polarizers P1 and P2 are used to filter the polarizations coming to beamsplitter BS2, which combines the paths resulting to a double-peaked frequency spectrum. As the recombined photon goes through polarizer P3, the photon's initial polarization state $\rho$ is fixed. The relative amplitude $A$ between the Gaussian peaks is controlled by rotating polarizers P1 and P2. Then, the photon arrives at the measurement stage where it first goes through the quartz plate combination (QPs) corresponding to the unitary coupling $U$, where the probe (polarization) and the system (frequency) interact. After the interaction, the photon goes through a combination of a quarter-wave plate (QWP), half-wave plate (HWP), and polarizer P4 before finally arriving at a single photon detector D, which are used together to perform full polarization state tomography to extract $\Phi(\rho)$.
  }
  \label{experimental_setup}
\end{figure*}

Once the other parameters of the Gaussian distributions are fixed, the convex coefficient  $A$ determines directly whether the polarization dynamics is Markovian or non-Markovian in the following way: If we first restrict to $A\in[0,1/2]$, $h(A)$ is a monotonically increasing function in $A$. This means, that there exists $A_{crit}\in [0,1/2]$ such that the dynamics is Markovian for all $A<A_{crit}$ and non-Markovian for all $A \in [A_{crit},1/2]$. On the other hand, if $A\in[1/2,1]$, $h(A)$ is a monotonically decreasing function in $A$ and the opposite holds. Thus, we conclude that the dynamics is non-Markovian if $A \in [A_{crit} , 1 - A_{crit} ]$ and Markovian otherwise.

Motivated by the analysis of pure dephasing channels in \cite{tukiainen,lyyrasiltanen2020}, we choose the initial probe states of the polarization system as $\rho_1 = \rho_2 = \rho_3 =  \ket{+}\bra{+}$. After the polarization has interacted with the frequency for an unknown time, we perform tomographic measurements on the evolved states $\Phi_1(\rho_1)$, $\Phi_2(\rho_2)$, and $\Phi_3(\rho_3)$ and calculate the $\alpha$-fidelities $F_{\alpha} \big(\Phi_1(\rho_1), \Phi_2(\rho_2)\big)$ and  $F_{\alpha} \big(\Phi_1(\rho_1), \Phi_3(\rho_3)\big)$. As $F_{\alpha} (\rho, \rho) = 1$ and $D_{tr} (\rho, \rho) = 0$ for any state $\rho$, Equations \eqref{ainequalityfull}, and \eqref{trainequality} with the measurement data give us simplified bounds for the amplitude parameter as
\begin{align}\label{ainequality3}
\begin{aligned}
1 - F_{\alpha_3} \big( \Phi_1(\rho), \Phi_3(\rho) \big)^{1/\alpha_3} \leq 
 A 
\leq  F_{\alpha_2} \big( \Phi_1(\rho), \Phi_2(\rho) \big)^{1/\alpha_2}\,,\\
D_{tr} \big( \Phi_1( \rho ), \Phi_3( \rho ) \big) 
 \leq   A  \leq 
1 - D_{tr} \big( \Phi_1( \rho ), \Phi_2( \rho ) \big)\,.
\end{aligned}
\end{align}
If any of the experimentally dertermined upper bounds is below $A_{crit}$, we immediately know that the polarization dynamics is Markovian, and similarly if any of the lower bounds is above $1 - A_{crit}$. On the other hand, if any of the lower bounds is larger than $A_{crit}$ and any upper bound is smaller than $1-A_{crit}$, the dynamical map of the polarization qubit is non-Markovian. The Markovian and non-Markovian regions of the $A$ parameter are illustrated in Fig.~\ref{acrit} \textbf{a)} . 

In Fig.~\ref{acrit} \textbf{b)}, we have numerically calculated $A_{crit}$ from the non-Markovianity condition Eq.~\eqref{nmcondforA} as a function of the ratio between the other free parameters $\Delta\eta:=\Delta\mu / \sigma$.
The numerical analysis 
suggests that whenever $\Delta\eta > 0$, there exists a non-empty interval $[A_{crit}, 1 - A_{crit}]$ such that the polarization dynamics is non-Markovian for all $A$ within the interval. 
%
%
It seems that when $\Delta\eta$ is large, even small values of $A$ produce non-Markovian dynamics. The function fitted to Eq.~\eqref{nmcondforA} illustrates this well. The non-Markovian region $[A_{crit},1-A_{crit}]$ is given by
\begin{equation}
A_{crit}(\Delta\eta) = 0.0885553 e^{-0.0870419 \Delta \eta ^2} + \frac{0.411445}{0.0845395 \Delta \eta ^2+1}\,.
\label{fit}
\end{equation}
Thus, $A_{crit}$ decreases monotonically as a function of $\Delta\eta$ and consequently, the larger $|\mu_2-\mu_1|$ (or the smaller $\sigma$), the larger the non-Markovian set of $A$ values $[A_{crit},1-A_{crit}]$, and vice versa.

\begin{figure*}[t]
\begin{minipage}[t!]{0.9\textwidth}
\centering
  \includegraphics[width=0.8\textwidth]{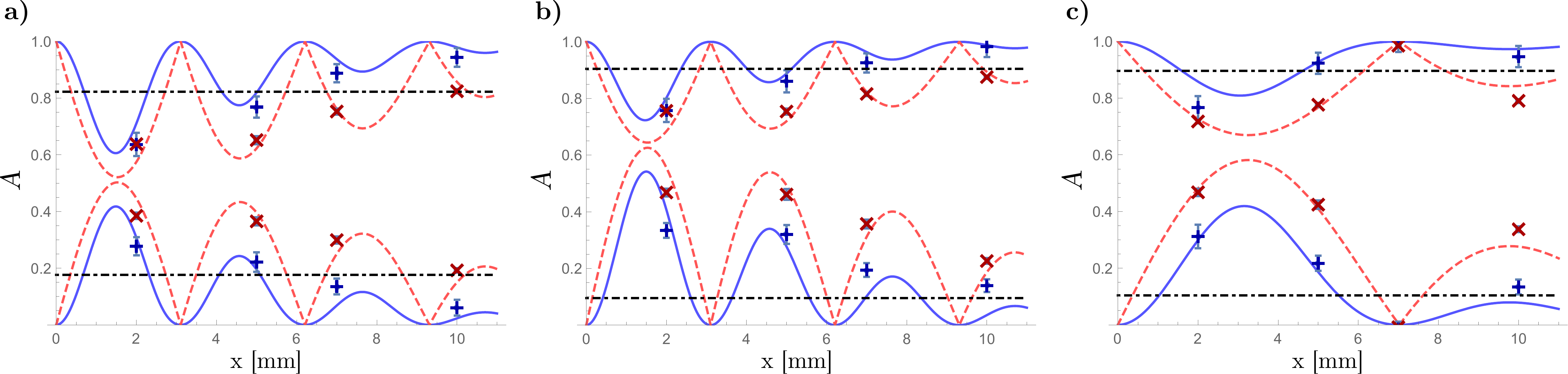}
\end{minipage}
  \caption{
  \textbf{The probed bounds for $A_{crit}$ and the convex coefficient $A$.} 
  \textbf{a)}: $\lambda_1 = 810$ nm, $\lambda_2 = 830$ nm, $A = 0.5122$. 
  \textbf{b)}: $\lambda_1 = 810$ nm, $\lambda_2 = 830$ nm, $A = 0.6377$. 
  \textbf{c)}: $\lambda_1 = 810$ nm, $\lambda_2 = 820$ nm, $A = 0.6377$. 
  The x axis is the thickness of the quartz plate combination. The blue crosses are the lower and upper bounds obtained from the $\alpha$-fidelity and the red slanted crosses are the bounds obtained from the trace distance. The blue solid and red dashed lines are the theoretical predictions for the bounds of the matching color, calculated taking into account dispersion in quartz. The black dashdotted lines are the values of $A_{crit}$ and $1 - A_{crit}$, limiting the non-Markovian region. The $A_{crit}$ values were obtained by probing of $\Delta \mu / \sigma$ with Eq.~\eqref{etabounds}. 
  The error bars are due to the photon-counting statistics, and they are standard deviations of the bound values calculated by the Monte Carlo method.
  }
  \label{resultset1_2_3}
\end{figure*}

%
Recently, a method for probing the 
lower bound for $\Delta\eta$ 
was experimentally implemented in the cases of unknown coupling between the system (frequency) and probe (polarization) \cite{lyyrasiltanen2020}. For our notation, the probed bound is given by
\begin{equation}\label{etabounds}
\Delta\eta \ge 
\sqrt{ \frac{ 2 \ln \left[ F_\alpha \left( \Phi_2 (\rho_2), \Phi_3 (\rho_3) \right) / F_\alpha \left( \rho_2, \rho_3 \right) \right] }{ \alpha(\alpha - 1) } }\,.
\end{equation}
Since $A_{crit}$ decreases monotonically in $\Delta\eta$, probing a lower bound for $\Delta\eta$ and inserting it in Eq.~\eqref{fit} would give us a pessimistic upper bound $\tilde A_{crit}$ for $A_{crit}$. If our measured bounds of $A$ are between $\tilde A_{crit}$ and $1 - \tilde A_{crit}$, then $A$ is guaranteed to be between the actual values $A_{crit}$ and $1 - A_{crit}$. Thus, the unknown coupling probing protocol can be used to extract appropriate bounds for each parameter to confirm that the dynamics is non-Markovian.

In the Appendix we analyze the limitations of using multiple quartz plates in the same orientation as system-probe coupling to verify that the probe dynamics is Markovian at all times $t>0$, or in other words, the global Markovianity of the probe dynamics. We conclude that such system-probe coupling always fails in that task. Even though that coupling cannot be used to determine the global Markovianity, we may use the same approach of unknown couplings as in \cite{lyyrasiltanen2020} and choose the rotation angles of each plate randomly. In \cite{lyyrasiltanen2020} random rotation angles improved the precision of the probing. In such a situation, the analysis of the measurement data is exactly the same, as one needs to just use the measured probe states in Eqs.~\eqref{ainequality3} and \eqref{etabounds} to extract the bounds for $A$ and $\Delta\eta$, respectively.

Contrary to the analysis of global Markovianity, the polarization's global non-Markovianity and Markovian time intervals can be conclusively determined with quartz plates in the same orientation, as we will see in section \ref{probing_results}. Next, we present our quantum optical experimental setup and use it to obtain the bounds for the parameters $A$ and $\Delta \eta$ and eventually determine if the polarization dynamics is non-Markovian.

\section{The experimental setup}

Our experimental setup is presented in Fig.~\ref{experimental_setup}. First, a pair of photons with wide frequency spectra is produced in a photon source (PS) by spontaneous parametric down-conversion process when a type I beta-barium borate crystal is pumped with a tightly focused continuous wave laser at the wavelength 405 nm. One of the photons, the idler, is guided directly to a single photon detector D which sends a trigger to the coincidence counting electronics to expect the signal photon at the other detector. 

The signal photon arrives at a beamsplitter (BS1), which turns the photon into a spatial superposition of branch 1 and branch 2. In branches 1 and 2, the photon passes through the interference filters IF1 and IF2 with FWHM of 3 nm, respectively. The transmission bands of IF1 and IF2 have different central frequencies $\mu_1$ and $\mu_2$, which filters the frequency spectrum in each branch to the Gaussian distribution $G_1$ and $G_2$. IF2 can be also tilted, and thus the distance of the Gaussians $\Delta\mu$ can be adjusted. 

In the branches, the photon goes through the polarizers P1 and P2 placed in rotation stations. Then, the branches are recombined with another beamsplitter (BS2), after which the photon passes through a third polarizer that prepares the initial polarization states $\rho_1 = \rho_2 = \rho_3 = \ket{+}\bra{+}$. Together with the polarizers in branches 1 and 2, the third polarizer P3 controls the amplitudes of the Gaussians $G_1$ and $G_2$ by dimming the Gaussian from each branch. The dimming of each Gaussian, and thus the value of the amplitude parameter $A$, is determined by the relative rotation angles between the branch polarizers P1 and P2 and the third polarizer, giving us high control of the state of interest $\xi_1$. Each of the branches can also be independently blocked, resulting to the reference states $\xi_2$ and $\xi_3$.

After the polarizer P3, the photon goes through a combination of quartz plates (QPs). In the quartz plates, the system (frequency) and probe (polarization) are coupled by unitary $U$, causing the probe to experience dephasing dynamics described by Eq.~\eqref{dephasing}. In the experiment, the interaction time is given by the thickness of the used QP combination. 

Once the interaction ends, the photon goes through a combination of a quarter-wave plate (QWP), half-wave plate (HWP), and polarizer P4, after which it is guided into a single photon detector. The waveplate combination and the polarizer P4 are used to perform tomography for the polarization qubit to obtain the evolved probe states $\Phi_1 (\rho)$, $\Phi_2 (\rho)$, and  $\Phi_3 (\rho)$.\\

\section{Measurement results}\label{probing_results}

\subsection{Global non-Markovianity}\label{global_n-m}

\begin{figure}[t]
\centering
  \includegraphics[width=1.0\linewidth]{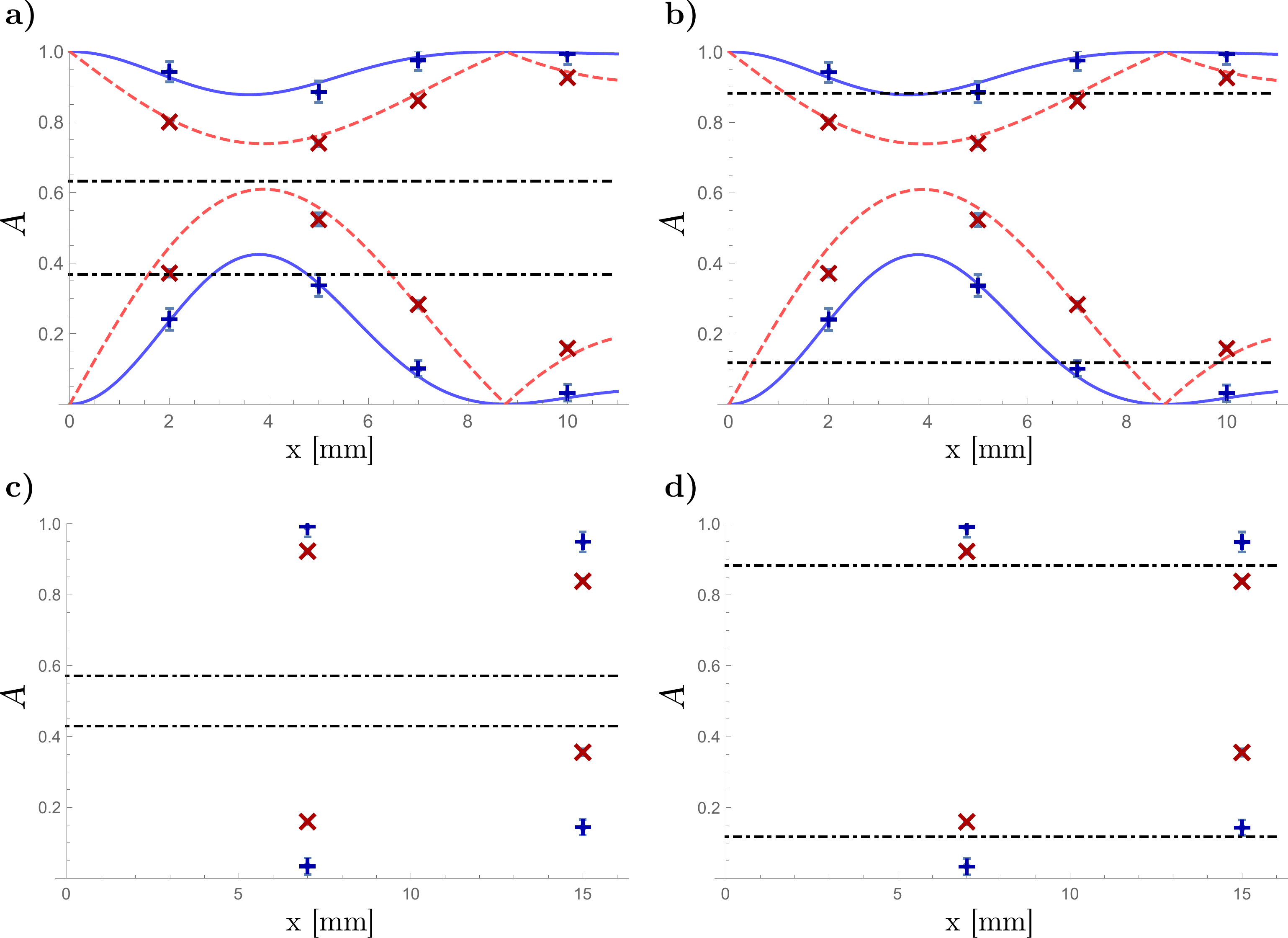}
  \caption{
  \textbf{The probed bounds for the convex coefficient $A$.}  \textbf{a) - d)}: $\lambda_1 = 810$ nm, $\lambda_2 = 818$ nm, $A = 0.7$. The x axis is the thickness of the quartz plate combination. The blue crosses are the lower and upper bounds obtained from the $\alpha$-fidelity and the red slanted crosses are the bounds obtained from the trace distance. The blue solid and red dashed lines are the theoretical predictions for the bounds of the matching color, calculated taking into account dispersion in quartz. The black dashdotted lines are the values of $A_{crit}$ and $1 - A_{crit}$, limiting the non-Markovian region. In \textbf{a)} and \textbf{b)} the fast axes of all the quartz plates were aligned. In \textbf{c)} and \textbf{d)} the rotation angle of each quartz plate was chosen randomly. In \textbf{a)} and \textbf{c)} the $A_{crit}$ values were obtained by probing of $\Delta \mu / \sigma$ with Eq.~\eqref{etabounds}. 
  In \textbf{b)} and \textbf{d)} the known value of $A_{crit}$ was used.
     The error bars are due to the photon-counting statistics, and they are standard deviations of the bound values calculated by the Monte Carlo method.
  }
  \label{resultset5_6}
\end{figure}

We present our experimental results in Figs.~\ref{resultset1_2_3}-\ref{resultset5_6}. 
In each case, we have used different choices for the central frequencies $\mu_k = c/\lambda_k$ of the Gaussians or the convex coefficient $A$, but the standard deviation $\sigma$ of the Gaussians is kept fixed. For each choice of the parameters, we let the probe (polarization) interact with the system (frequency) and performed full tomography of the evolved probes for each initial system state. This was repeated for multiple system-probe couplings, corresponding to different thicknesses of QP combinations, as illustrated by the horizontal axes of the figures. We note here that the thicknesses are shown only to compare the measurement data with the theoretical predictions and they were not used to make any conclusions about non-Markovianity.

For each coupling, the probed lower and upper bounds for $A$ were obtained by using the results of evolved probe state tomographies $\Phi_1 (\rho)$, $\Phi_2 (\rho)$, $\Phi_3 (\rho)$ in Eq.~\eqref{ainequality3} corresponding to both the trace distance (red slanted crosses) and the $\alpha$-fidelity (blue crosses). For the $\alpha$-fidelity bounds, we used $\alpha_2 = \alpha_3 = 1/2$, since numerical tests showed it to result to tightest probed bounds for $A$. 

In the measurements presented in Figs.~\ref{resultset1_2_3}, \ref{resultset5_6} \textbf{a)}, and \ref{resultset5_6} \textbf{c)}
, we used the method of \cite{lyyrasiltanen2020} to extract lower bounds for $ \Delta\mu /\sigma$, which we used to obtain upper bounds for $A_{crit}$ corresponding to Eq.~\eqref{fit}. For each of these cases, we used the same measurement data as when probing bounds of $A$, but the $\alpha$ values were optimized independently for each used coupling to obtain the tightest bound, as done in \cite{lyyrasiltanen2020}. Like in \cite{lyyrasiltanen2020}, the optimal $\alpha$ value varied and it was most commonly near $\alpha = 1$.


The panels in Fig.~\ref{resultset1_2_3} show that our probing approach managed to extract tight enough bounds for the unknown parameters $A$ and $ \Delta\mu /\sigma$ to verify the non-Markovianity of the dynamics: in each case, at least for one system-probe coupling the probed lower and upper bounds are between the probed $A_{crit}$ and $1 - A_{crit}$ which define the non-Markovian region. Thus in each case we could determine that the dynamics was non-Markovian from the outcome of a single snapshot measurement. 

On the other hand, for some couplings the result is inconclusive, as either the probed lower bound is below $A_{crit}$ or the probed upper bound is above $1 - A_{crit}$. We note that in this case the bounds obtained using trace distance led to a lot tighter bounds than the ones derived from the $\alpha$-fidelities. 
The probed upper bound of $A_{crit}$ in Fig.~\ref{resultset5_6} \textbf{a)} 
was too large leading to inconclusive result in verifying non-Markovianity. Figure \ref{resultset5_6} \textbf{b)} 
shows that when the exact value of $A_{crit}$ was known, the probed lower and upper bounds of $A$ were tight enough to verify the non-Markovianity of the polarization dynamics.

For the measurements presented in Figs.~\ref{resultset5_6} \textbf{c)} and \textbf{d)}
, we used a completely unknown system-probe coupling by fixing the quartz plates in randomly chosen rotation angles, like done earlier in \cite{lyyrasiltanen2020}. As above, we used this unknown coupling to probe if the probe's dynamics would be non-Markovian if all the quartz plates in the combination were aligned.  In Fig.~\ref{resultset5_6} \textbf{c)}
, we used the measurement data to extract bounds for both $A$ and $A_{crit}$. The data shows that as in the case of Fig.~\ref{resultset5_6} \textbf{a)}, 
 this measurement led to inclonclusive verification of non-Markovianity. In Fig.~\ref{resultset5_6} \textbf{d)} 
 the exact value of $A_{crit}$ was assumed as known and we see that this time the measurement data led to conclusive verification of non-Markovianity. This means that quantum probing measurements with unknown system-probe interactions can be used to limit the convex coefficient to a non-trivial interval which in turn can be exploited to make conclusions on the characteristics of the BLP non-Markovianity.

%
%

To summarize, in each measurement we succesfully verified the non-Markovianity of the polarization dynamics by probing $A$ at a single unknown interaction time. Additionally, in the measurements presented in Fig.~\ref{resultset1_2_3} 
we managed to exploit the same measurement data to probe small enough upper bounds for $A_{crit}$ that the non-Markovianity could be verified without assuming any of the parameters in the frequency states.

We emphasize that in order to make conclusions of the non-Markovianity, we only need to measure the probe system evolved with the map of interest only at one unknown time and compare it to our reference maps with the same unknown interaction time. The protocol itself does not require any information on the actual interaction time, or more generally, any knowledge of the system-probe coupling $U$ as demonstrated by the results in Fig.~\ref{resultset5_6} \textbf{d)}
. Even though the general form of the dephasing dynamics in this model is well-known \cite{liu2011,liu2018}, measuring the evolved polarization state at an unknown interaction time cannot tell anything about the Markovianity or non-Markovianity of the qubit dynamics, as many choices of $A$ can lead to the same value $|\kappa(t)|$, corresponding to the distinguishability of the optimal pair of states, even if the other parameters are fixed. 

\begin{figure}[H]
\centering
\includegraphics[width=1.0\linewidth]{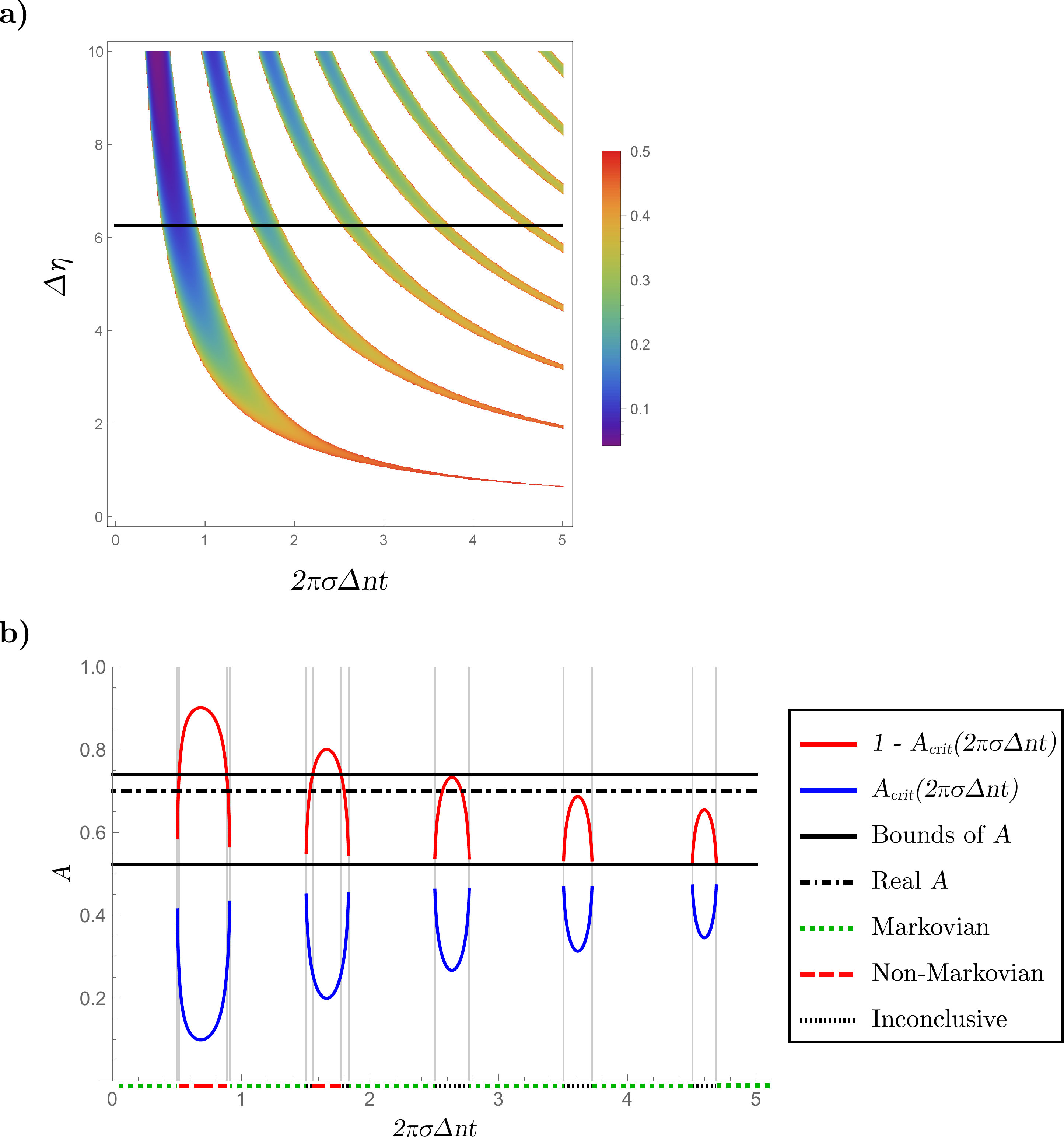}
\caption{\textbf{Probing the Markovian and non-Markovian time intervals.} (Color online) \textbf{a):} The time and $\Delta\eta$ dependence of the critical amplitude $A_{crit}$. The dynamics is Markovian in the white regions for all $A$. The horizontal line shows the $\Delta\eta$ in panel \textbf{b)} where we have fixed $\lambda_1 = 810$ nm, $\lambda_2 = 818$ nm, $A = 0.7$. In \textbf{b)}, the lower and upper bound of $A$ are the tightest bounds in Fig.~\ref{resultset5_6} \textbf{a)}. Since the probed lower and upper bounds are in $[A_{crit},1-A_{crit}]$ during the time intervals marked with red dashed lines, we verify that the dynamics at those times is non-Markovian. During the intervals marked with green dotted lines neither the lower nor upper bound is in $[A_{crit},1-A_{crit}]$, and we verify that the dynamics is Markovian at these time intervals. Vertical gray lines are guide for the eye. The x axis is the rescaled interaction time inside the quartz plate combination. Here,  $\Delta\eta$ is assumed as known.}
\label{acritt}
\end{figure}

\subsection{Markovian and non-Markovian time intervals}\label{n-m_intervals}

Finally, we show how our probing results can be exploited to identify the time intervals where the dynamics is guaranteed to be Markovian or non-Markovian. Figure \ref{acritt} \textbf{a)} shows the time dependence of $A_{crit}$ for different values of $\Delta\eta$, which was determined numerically with Eq.~\eqref{nmcondforA}. Inside the white areas, the dynamics is Markovian for all $A$ in $[0,1]$. We see that as $\Delta\eta$ decreases, the first non-Markovian period appears later. The plot also shows that $A_{crit}$ is always smallest during the first non-Markovian period and larger $\Delta\eta$ leads to smaller $A_{crit}$. These observations are in good agreement with the non-Markovian behavior of the frequency-polarization model, since $\Delta\mu \propto \Delta\eta$ gives rise to the revivals and $\sigma \propto \Delta\eta^{-1}$ corresponds to the damping rate of the trace distance \cite{liu2011,liu2018}. The horizontal black line highlights the fixed value of $\Delta\eta$ in Figs.~\ref{resultset5_6} and \ref{acritt} \textbf{b)}.

In Fig.~\ref{acritt} \textbf{b)} we plot the time dependence of $A_{crit}$ for the value of $\Delta\eta$ in the measurements of Fig.~\ref{resultset5_6}. As in Fig.~\ref{acritt} \textbf{a)}, we see time intervals where $A_{crit}$ is not defined, corresponding to times when there does not exist such $A \in [0,1]$ that would satisfy Eq.~\eqref{nmcondforA}. Thus we know that the dynamics on all those intervals is Markovian, marked with green dotted x axis. The blue and red solid curves limit the non-Markovian region $[A_{crit},1-A_{crit}]$ and the black dashdotted line marks the real value of $A$ in the experiment. 

We see that for the rescaled interaction time $2\pi\sigma\Delta n t$ in $[0,3]$ the real value of $A$ is between $A_{crit}$ and $1-A_{crit}$ on three time intervals, meaning that the probe dynamics is really non-Markovian at those times. Looking at the black solid lines, corresponding to the tightest probed lower and upper bounds for $A$ in Fig.~\ref{resultset5_6} \textbf{a)}, we see that for the first two non-Markovian time intervals  the lower and upper bound are on the interval $[A_{crit},1-A_{crit}]$. This means that our probed bounds of $A$ combined with our analysis on the model's non-Markovianity lets us verify that the probe dynamics was non-Markovian on those intervals. For these confirmed non-Markovian intervals the x axis is marked with red dashing. 

For the three potentially non-Markovian intervals for $2\pi \sigma \Delta n t\in[2,5]$, the probed upper bound is above $1-A_{crit}$ while the probed lower bound is between $A_{crit}$ and $1-A_{crit}$, so our probing measurement leads to inconclusive result on the non-Markovianity on those intervals, marked with black dotted x axis. We note also the very small inconclusive intervals around the two confirmed non-Markovian intervals.

Here, we probed the Markovian and non-Markovian time intervals only for the measurement in Fig.~\ref{resultset5_6}. The same analysis can be directly applied also to the rest of our measurements.

\section{Conclusions and Outlook}

In this paper we applied the generalized data processing inequalities of $\alpha$-fidelities and trace distance to construct a quantum measurement strategy for probing lower and upper bounds for the convex coefficients in mixtures of commuting states. The measurement strategy does not require any knowledge of the used system-probe coupling and it can be directly applied with no modifications if the coupling is changed. We first discussed briefly some possible applications. Then, we explored in detail a specific task, namely the verification of the probe's non-Markovianity, a useful property for certain quantum information protocols \cite{panalgorithm,karlssonEPL,liu2020,laine2014, huelgarivasplenio2012, Xiang2014,utagi}, by snapshot probing measurement at an unknown time and with a completely unknown system-probe coupling. 

We showed that when a single photon's polarization interacts with the photon's double peaked Gaussian frequency spectrum in quartz, the non-Markovian behavior of the polarization dynamics is fully contained in an intact and well-defined area in the two-dimensional $(A,\Delta\mu/\sigma)$ parameter space. Here, $A$ is the convex coefficient in the mixture of the Gaussian peaks, $\Delta\mu$ is the difference of their central frequencies, and $\sigma$ is their standard deviation. We applied our strategy in probing lower and upper bounds for the convex coefficient $A$ and exploited the same measurement data in the recently proposed strategy to probe a lower bound for $\Delta\mu/\sigma$ \cite{lyyrasiltanen2020}. 
The probing strategies were implemented in an all-optical single photon experiment where we were able to restrict the unknown parameters $A$ and $\Delta\mu/\sigma$ within an area where the non-Markovianity of the polarization dynamics is guaranteed. By assuming $\Delta\mu/\sigma$ as known, we applied our probing results of $A$ to identify the Markovian and non-Markovian time intervals of the polarization dynamics. Thus, our results illustrate that quantum probing measurements with unknown system-probe couplings can be constructed and implemented to obtain useful qualitative information on the characteristics of the probe's dynamical map.

Even though we concentrated here in non-Markovianity in terms of revivals of trace distance, in our case of dephasing dynamics, these results apply directly to multiple other definitions of non-Markovianity as well, namely violation of CP-divisibility, Bloch volume oscillations, and increases of $l_1$ coherence norm \cite{teittinen2018}. In the dephasing dynamics, revivals of trace distance also imply that the quantum speed limit bound is not reached \cite{deffnerlutz,xuspeed,teittinen2019}. Thus, our results can be directly used to conclude that the probe dynamics is not on its fastest trajectory when the dynamics is verified as non-Markovian.


Our verification strategy is based on quantum probing measurements. It has been recently shown that using probes initially entangled with an ancillary system can achieve higher precision in quantum probing \cite{girolami}. However, adding an ancillary system increases the total probe-ancilla Hilbert space dimension, which makes the required tomography more demanding. Recently, experimentally estimating the fidelity between two-photon polarization states was shown to be more efficient than full tomography \cite{teiko,lyyra}. The results and experimental implementation can be directly generalized to $\alpha$-fidelities. Future studies will show if using entangled ancillary polarization system can be exploited to obtain better sensitivity in verification of non-Markovianity without the need of increasing the amount of measurements significantly.

\section{Acknowledgments}

H.~L. acknowledges the fruitful discussions with Erkka Haapasalo and Juha-Pekka Pellonpää. O.S. acknowledges the financial support from Magnus Ehrnrooth Foundation.
\vspace{1cm}

\appendix

\renewcommand{\theequation}{A\arabic{equation}}

\setcounter{equation}{0}

\section*{Appendix: Analysis of probing global Markovianity}


\begin{figure}[t!]
\centering
\includegraphics[width=0.9\linewidth]{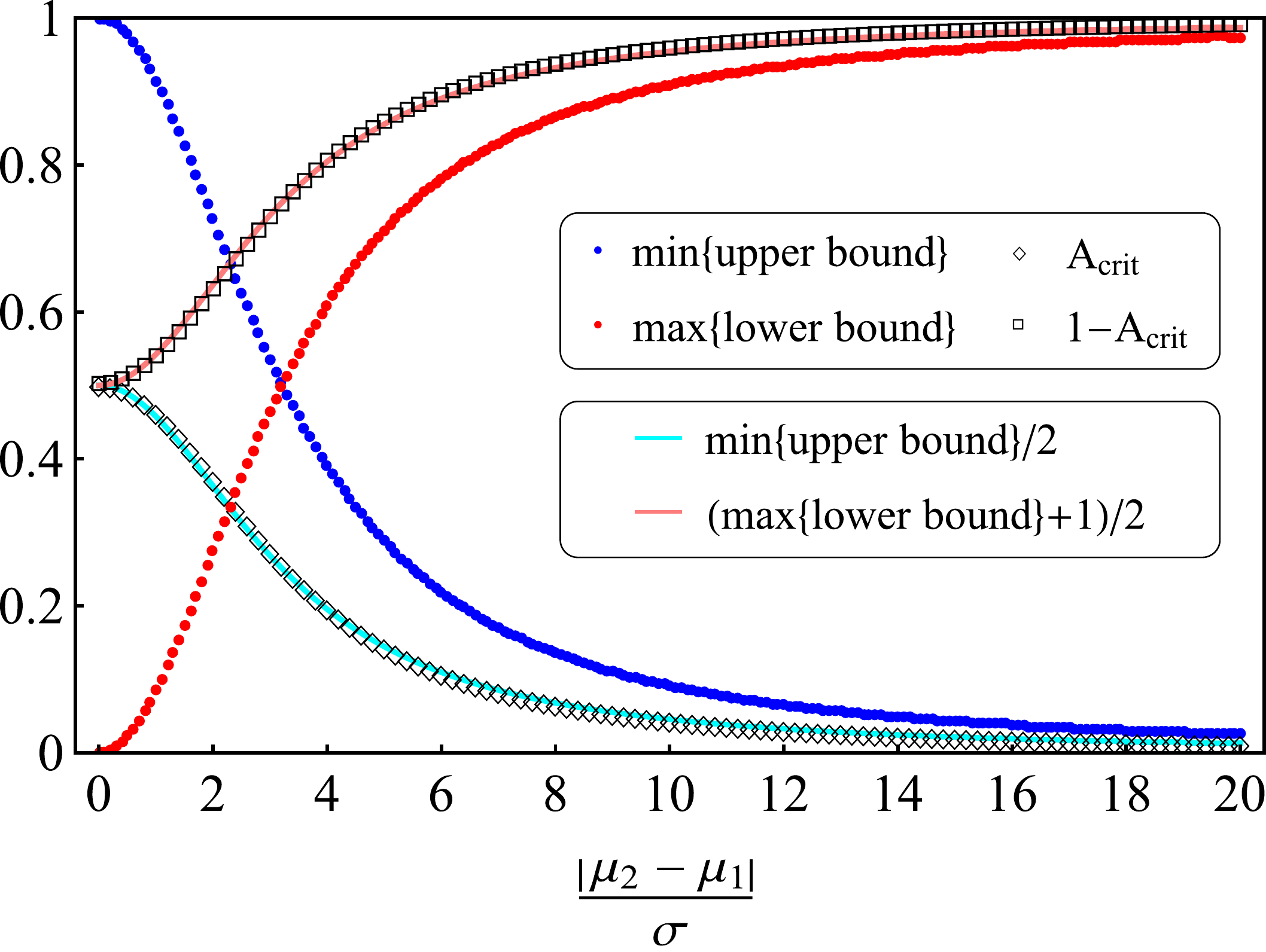}
\caption{(Color online) The tightest possible upper and lower bounds probed with the $\alpha$-fidelity (blue and red dots) optimized w.r.t. time, $\alpha$'s, and $A$ independently, $A_{crit}$ and $1-A_{crit}$ (black diamonds and squares), and the corresponding fitted curves (cyan and pink solid lines) as functions of $| \mu_2-\mu_1 | / \sigma$.}
\label{numerical_plots}
\end{figure}

Here we discuss the limitations of our probing measurements when using aligned quartz plates as the unitary coupling in verifying Markovianity of the polarization dynamics. To conclude whether the dynamics is Markovian, the experimentally obtained upper bound for the convex coefficient must be less than the convex coefficient's critical value, i.e.,
\begin{equation}
F_{\alpha_2}\big(\Phi_1(\rho),\Phi_2(\rho)\big)^{1/\alpha_2}<A_{crit}.
\label{concl_Mark1}
\end{equation}
Alternatively, the experimentally obtained lower bound must satisfy
\begin{equation}
1-F_{\alpha_3}\big(\Phi_1(\rho),\Phi_3(\rho)\big)^{1/\alpha_3}>1-A_{crit}.
\label{concl_Mark2}
\end{equation}

Numerical analysis shows that $F_{\alpha_2}\big(\Phi_1(\rho),\Phi_2(\rho)\big)^{1/\alpha_2}$ grows with $\alpha_2 \in [1/2,1)$ and $A \in [0,1]$. Since smaller $F_{\alpha_2}\big(\Phi_1(\rho),\Phi_2(\rho)\big)^{1/\alpha_2}$ and $F_{\alpha_3}\big(\Phi_1(\rho),\Phi_3(\rho)\big)^{1/\alpha_3}$ lead to tighter bounds in Eq.~\eqref{ainequality3}, we use $\alpha_2 = \alpha_3 = 1/2$ in the measurement data analysis when probing the bounds of $A$. To see if Eq.~\eqref{concl_Mark1} is ever satisfied, we may fix $\alpha_2=1/2$, $A=0$, and only consider $\min_t \Big\{F_{1/2}\big(\Phi_1(\rho),\Phi_2(\rho)\big)^2\big\vert_{A=0}\Big\}$, where the minimum is taken over all system-probe interaction times $t$. We have plotted the values of $\min_t \Big\{F_{1/2}\big(\Phi_1(\rho),\Phi_2(\rho)\big)^2\big\vert_{A=0}\Big\}$ as a function of $\Delta\eta$ in Fig.~\ref{numerical_plots}, where the values of $A_{crit}$ are shown for comparison. We notice that the values of $A_{crit}$ are in excellent agreement with $\min_t\Big\{F_{1/2}\big(\Phi_1(\rho),\Phi_2(\rho)\big)^2\big\vert_{A=0}\Big\}/2$ for all choices of $\Delta\eta$. Thus, we can estimate that
\begin{widetext}
\begin{equation}
F_{\alpha_2}\big(\Phi_1(\rho),\Phi_2(\rho)\big)^{1/{\alpha_2}}
\geq\min_t \Big\{F_{1/2}\big(\Phi_1(\rho),\Phi_2(\rho)\big)^2\big\vert_{A=0}\Big\}
\approx2A_{crit}>A_{crit}.
\label{estimation1}
\end{equation}
A similar analysis for the lower bound results in the estimation
\begin{equation}
1-F_{\alpha_3}\big(\Phi_1(\rho),\Phi_3(\rho)\big)^{1/{\alpha_3}}\leq\max_t \Big\{1-F_{1/2}\big(\Phi_1(\rho),\Phi_3(\rho)\big)^2\big\vert_{A=1}\Big\}\approx1-2A_{crit}<1-A_{crit}.
\label{estimation2}
\end{equation}
Combining Eqs.~\eqref{estimation1} and~\eqref{estimation2}, we conclude that, for any choice of parameters $A$, $\mu_1$, $\mu_2$, $\sigma$, $\alpha_2$, $\alpha_3$, and $t$, we get
\begin{equation}
[1-F_{\alpha_3}\big(\Phi_1(\rho),\Phi_3(\rho)\big)^{1/{\alpha_3}},F_{\alpha_2}\big(\Phi_1(\rho),\Phi_2(\rho)\big)^{1/{\alpha_2}}]\cap[A_{crit},1-A_{crit}]\neq\emptyset,
\label{conclusion}
\end{equation}
\end{widetext}
meaning that the bounds obtained by the $\alpha$-fidelity cannot be used to confirm the global Markovianity of the qubit dynamics when the quartz plates are aligned in the same orientation.


We can derive the same result for the bounds obtained by the trace distance by using the estimation $A_{crit}\approx\frac{1}{2}\min_t \Big\{F_{1/2}\big(\Phi_1(\rho),\Phi_2(\rho)\big)^2\big\vert_{A=0}\Big\}$. Below, we show that the experimentally obtained upper bound satisfies $1-D_{tr}\big(\Phi_1(\rho),\Phi_2(\rho)\big)\geq A_{crit}$ and thus cannot confirm Markovianity. Using the shorthand notation $\tau=2\pi\sigma\Delta nt$, we get
\begin{widetext}
\begin{align}
\Big[e^{-\frac{1}{2}\tau^2}\sin^2\Big(\frac{\Delta\eta\tau}{2}\Big)-e^{\frac{1}{2}\tau^2}\Big]^2&\geq 0 ~~\forall~\tau \ge 0\\
\Leftrightarrow 2\sin^2\Big(\frac{\Delta\eta\tau}{2}\Big)&\leq e^{\tau^2}+e^{-\tau^2}\sin^4\Big(\frac{\Delta\eta\tau}{2}\Big) ~~\forall~\tau \ge 0\\
\Leftrightarrow 2-2\cos(\Delta\eta\tau)&\leq e^{\tau^2}-\cos(\Delta\eta\tau)+1+\frac{1}{4}e^{-\tau^2}\big[\cos(\Delta\eta\tau)-1\big]^2 ~~\forall~\tau \ge 0\\
\Leftrightarrow\sqrt{2-2\cos(\Delta\eta\tau)}&\leq e^{\frac{1}{2}\tau^2}-\frac{1}{2}e^{-\frac{1}{2}\tau^2}\big[\cos(\Delta\eta\tau)-1\big] ~~\forall~\tau \ge 0\\
\Leftrightarrow 1-\frac{1}{2}e^{-\frac{1}{2}\tau^2}\sqrt{2-2\cos(\Delta\eta\tau)}&\geq\frac{1}{2}\Big\{1+\frac{1}{2}e^{-\tau^2}\big[\cos(\Delta\eta\tau)-1\big]\Big\} ~~\forall~\tau \ge 0\\
\Rightarrow 1-\frac{1-A}{2}e^{-\frac{1}{2}\tau^2}\sqrt{2-2\cos(\Delta\eta\tau)}&\geq\frac{1}{2}\min_t \Big\{1+\frac{1}{2}e^{-\tau^2}\big[\cos(\Delta\eta\tau)-1\big]\Big\}~~\forall\,A\in[0,1]\\
\Leftrightarrow1-D_{tr}\big(\Phi_1(\rho),\Phi_2(\rho)\big)&\geq\frac{1}{2}\min_t \Big\{F_{1/2}\big(\Phi_1(\rho),\Phi_2(\rho)\big)^2\big\vert_{A=0}\Big\}\\
\Leftrightarrow 1-D_{tr}\big(\Phi_1(\rho),\Phi_2(\rho)\big)&\geq A_{crit}
\end{align}
\end{widetext}
which proves the claim. Similar calculation, using $A_{crit}\approx\frac{1}{2}\min\Big\{F_{1/2}\big(\Phi_1(\rho),\Phi_3(\rho)\big)^2\big\vert_{A=1}\Big\}$, holds for the lower bounds. 

Thus, we conclude that combinations of quartz plates in the same orientation cannot be used as the system-probe coupling to confirm the Markovianity of the probe dynamics. In this analysis, we concentrated on the global Markovianity, meaning that there are no revivals of the trace distance at any time $t\in [0,\infty)$. 

If instead we are more interested in local Markovianity, meaning monotonicity of trace distance on some finite interval $[t_1,t_2]$, the amount of trace distance revivals decreases - and consequently - the value of $A_{crit}$ increases. Thus, by restricting our interest to shorter intervals, we can confirm the Markovianity of the dynamics. In this approach, the same measurement data and Eq.~\eqref{ainequality3} can be used, but only the values of $A_{crit}$ should be calculated again according to the time interval when interpreting the results. Similar approach can be applied to cases where we are interested in the non-Markovianity at some specific time intervals. This was successfully implemented in Section \ref{n-m_intervals} of the main article.

\end{document}